\documentclass{PoS}

\usepackage{amsmath}
\usepackage{feynmp}

\title{Off-shell initial state effects and gauge invariance in the Drell-Yan process}

\ShortTitle{Off-shell initial states in the Drell-Yan process}

\author{\speaker{Maxim Nefedov}\thanks{Work supported in part by the Foundation for the Advancement of Theoretical Physics and Mathematics BASIS, grant No. 18-1-1-30-1 and Samara University Competitiveness Improvement Program under Task No. 3.5093.2017/8.9.}\\
        Samara National Research University,\\
        II Institute for Theoretical Physics, Hamburg University\\
        E-mail: \email{nefedovma@gmail.com}}

\author{Vladimir Saleev\\
        Samara National Research University\\
        E-mail: \email{saleev@samsu.ru}}

\abstract{The Helicity Structure Functions in the Drell-Yan process are discussed in a framework of Parton Reggeization Approach, which includes the transverse momentum of initial-state partons in a way compatible with QED gauge-invariance. Relationships with conventional Transverse Momentum Dependent Parton Model formalism are clarified.}

\FullConference{XXVII International Workshop on Deep-Inelastic Scattering and Related Subjects - DIS2019\\
		8-12 April, 2019\\
		Torino, Italy}

\begin{document}
\begin{fmffile}{fgraphs}

\section{Introduction}

   In the notation of Ref.~\cite{AMS} the differential cross-section of the Drell-Yan process of production of the lepton pair ($l^+l^-$) with transverse momentum ($q_T$), squared invariant mass ($Q^2=q^2$) in the collision of two non-polarized hadrons with center-of-mass energy ($\sqrt{S}$) can be written as:
\begin{eqnarray}
 \frac{d\sigma}{dx_A dx_B d^2{\bf q}_T d\Omega} &=& \frac{\alpha^2}{4Q^2} \Bigl[ F_{UU}^{(1)}\cdot
 \left( 1+\cos^2 \theta \right) + F_{UU}^{(2)}\cdot\left( 1-\cos^2 \theta \right)+\nonumber\\
  &+&F_{UU}^{(\cos \phi)} \cdot \sin(2\theta)\cos\phi +F_{UU}^{(\cos 2\phi)} \cdot \sin^2\theta\cos(2\phi) \Bigr], \label{eq:SFs-def}
\end{eqnarray}
were angles $\theta$ and $\phi$ define the direction of momentum of $l^+$ in the Collins-Soper frame~\cite{CS-frame}, $F_{UU}^{(1,2,\ldots)}(x_A,x_B,q_T)$ are the Helicity Structure Functions (HSFs) and $x_{A,B}=Q e^{\pm Y}/\sqrt{S}$.  

  In the present contribution we will present a QED gauge-invariant version of Transverse Momentum Dependent(TMD) factorization for the HSFs, based on the Parton Reggeization Approach (PRA)~\cite{DY-paper-1, DY-paper-2}. In the traditional TMD factorization, which for the purposes of this paper we will call the TMD Parton Model(TMD PM), see e.g.~\cite{AMS, CollinsQCD}, hadronic tensor does not satisfy the Ward identity of QED. PRA is a particular, physically motivated proposal for the $O(q_T/Q)$ corrections to the usual TMD hadronic tensor, which restore it's gauge-invariance. 

  Present contribution has the following structure. In the Sec.~\ref{sec:TMD-PM} we recall the notation of traditional TMD PM and in the Sec.~\ref{sec:PRA} we describe PRA and it's relationships with TMD PM and present some numerical results for HSFs.   

\section{TMD Parton Model}
\label{sec:TMD-PM}

  In the standard TMD PM approach~\cite{AMS, CollinsQCD}, based on a simple $q\bar{q}$-annihilation picture of the Drell-Yan process, the hadronic tensor is decomposed as follows:
\begin{eqnarray}
W_{\mu\nu}=W_{\mu\nu}^{\rm (TMD)}+Y_{\mu\nu}=\sum_{q,\bar{q}} \frac{e_q^2}{N_c} {\rm tr}\left[\gamma_\mu \Phi_{q}(q_1,P_1) \otimes_T \gamma_\nu \bar{\Phi}_{\bar{q}}(q_2,P_2)  \right] + Y_{\mu\nu}, \label{eq:TMD-PM}
\end{eqnarray}
where $f_1({\bf q}_{T1}) \otimes_T f_2({\bf q}_{T2})=\int d^2{\bf q}_{T1} d^2{\bf q}_{T2} \delta({\bf q}_T-{\bf q}_{T1}-{\bf q}_{T2}) f_1({\bf q}_{T1}) f_2({\bf q}_{T2})$ and four-momenta of quark($q_1$) and anti-quark($q_2$) are parametrized as $q^\mu_{1,2}=P^\mu_{1,2}x_{A,B}+q^\mu_{T1,2}$. The first term in Eq.~(\ref{eq:TMD-PM}) is a contribution of a leading power in $q_T/Q$ to a hadronic tensor, while all subleading contributions are supposed to be included to the $Y_{\mu\nu}$-term. Due to a large boost between hadron rest frame and hadronic center-of-mass frame, only terms proportional to $n_-^\mu=2P_1^\mu/\sqrt{S}$ contribute to the correlation function of quark fields $\Phi_q(q_1,P_1)$ at leading power, and it's Dirac structure can be parametrized as follows:
\begin{equation}
\Phi_q(q_1,P_1)=\frac{1}{2}\left[ \hat{n}_- f_1^{q}(x_1,q_{T1}) - i\sigma^{i-}\gamma_5 \frac{\epsilon_T^{ij} q^j_{T1}}{\Lambda} h_{1}^{\perp q}(x_1,q_{T1}) \right], \label{eq:Phi-TMD}
\end{equation} 
where, $f_1^{q}(x_1,q_{T1})$ is a number-density TMD Parton Distribution Function(PDF), $ h_{1}^{\perp q}(x_1,q_{T1})$ is a Boer-Mulders function~\cite{B-M}, $\Lambda$ is a scale of non-perturbative intrinsic transverse momentum of partons inside a hadron, which is typically taken to be $\Lambda\sim M$, and analogous decomposition holds for $\bar{\Phi}_{\bar{q}}$.

  The full hadronic tensor should satisfy Ward identity of QED:
\[
q^\mu W_{\mu\nu}=0,
\]
however, it is easy to verify, that for the first term in Eq.~(\ref{eq:TMD-PM}): $q^\mu W_{\mu\nu}^{\rm (TMD)}=O(q_T)$, so that the gauge-invariance is restored by some $O(q_T/Q)$ power-corrections from $Y_{\mu\nu}$. On the other hand, the $Y$-term is needed to describe the $q_T\gtrsim Q$-region, and phenomenologically we expect it to be well-approximated in this region by Collinear Parton Model calculation in the fixed-order in $\alpha_s(Q^2+q_T^2)$, while at $q_T\lesssim Q$ it should be negligible. Therefore, it is desirable to remove any genuinely non-perturbative contributions, associated with restoration of gauge-invariance of $W^{\rm (TMD)}$-term at $q_T\lesssim Q$, from the $Y$-term.       

  The problem of gauge-invariant definition of $W^{\rm (TMD)}$-term has been considered in Ref.~\cite{CollinsQCD} (Sec. 14.5.2). There it has been proposed to to put momenta of initial-state quarks on-shell: $q_{1,2}^2=0$, while retaining their transverse momenta and ``large'' light-cone momentum components. We call this approach -- Quasi-on-Shell Scheme (QOS). This scheme can be implemented in two ways, see Sec. 4 of our Ref.~\cite{DY-paper-1}, and resulting $Q(q_T/Q)$-corrections to the Hard-scattering Coefficients come-out to be totally different, depending on the prescription one chooses. Therefore one can not proceed with ad-hoc prescriptions and actually needs a physically-motivated ansatz for the power-supressed terms restoring the gauge-invariance of $W^{\rm (TMD)}$.
     
\section{Parton Reggeization Approach}
\label{sec:PRA}

  The gauge-invariange of the hadronic tensor holds because apart from the $t$-channel $q\bar{q}$ - annihilation (Parton Model) diagram, there exist other contributions to $p+p\to \gamma^\star + X$-amplitude, where photon is interacting directly with constituents of the colliding protons and beam-remnants. This contributions are beyond the scope of PM, but one can try to analyze them in model field theories and look for the limit when contributions of this kind also factorize, leading to some PM-like interpretation, independent on the details of above-mentioned interactions. 

  In fact, such factorization is well-known in the field of small-$x$ physics. It is proven in the Leading and Next-to-Leading Logarithmic Approximation w.r.t. resummation of $\log(1/x)$ in QCD~\cite{F-S, B-F-NLL}, that in the Multi-Regge limit $Q^2, q_T^2\ll S$ the universal vertex (Fadin-Sherman vertex) of production of virtual photon in an annihilation of Reggeized quark and antiquark factorizes-out from the amplitude:
\begin{equation}
\Gamma_\mu(q_1,q_2)=\gamma_\mu-\hat{q}_1\frac{n_\mu^-}{q_2^-}-\hat{q}_2 \frac{n_\mu^+}{q_1^+},\label{eq:F-S-vert}
\end{equation}
where $n_+^\mu=P_2^\mu/\sqrt{S}$. The vertex (\ref{eq:F-S-vert}) satisfies the Ward identity $(q_1+q_2)^\mu \Gamma_\mu(q_1,q_2)=0$.

  It is instructive to understand, how this factorization arises at tree level in a model field-theory, similar to that of Ref.~\cite{M-MFT}, where elementary spinorial ``proton'' fields ($p$) with electric charge $e_p$ and scalar ``spectator'' fields ($s$) with charge $e_s$ can interact with quarks with electric charge $e_q$. In this theory the following ``Drell-Yan'' process is possible:
\begin{equation}
\bar{p}(P_1)+p(P_2)\to \gamma^{\star}(q)+s(P_1')+s(P_2'), \label{eq:model-DY}
\end{equation}
and in the Multi-Regge limit: $S\gg Q^2\sim q_T^2$ one has 
\[
(P_1')^+\simeq P_1^+=\sqrt{S},\ (P_2')^-\simeq P_2^-=\sqrt{S},\ q_T\sim q^{\pm}\ll \sqrt{S}.
\]

  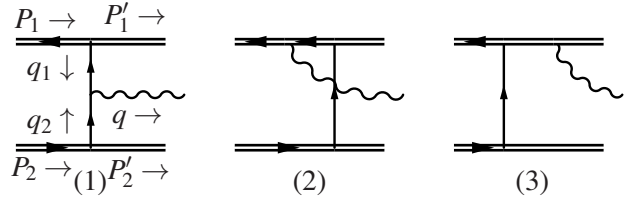
\begin{figure}[t]
  \begin{center}
\begin{tabular}{ccc}
\parbox{70pt}{\begin{fmfgraph*}(70,40)
\fmfset{curly_len}{1mm}
\fmfset{arrow_len}{2mm}
\fmfleft{l1,l2,l3}
\fmfright{r1,r2,r3}
\fmf{dbl_plain_arrow,label=$P_2\rightarrow\ \ \ $,label.side=right,label.dist=3.5pt}{l1,v1}
\fmf{dbl_plain,label=$\ \ \ P'_2\rightarrow$,label.side=right,label.dist=3.5pt}{v1,r1}
\fmf{dbl_plain_arrow,label=$P_1\rightarrow\ \ $,label.side=right,label.dist=3.5pt}{v3,l3}
\fmf{dbl_plain,label=$\ \ P'_1\rightarrow$,label.side=left,label.dist=3.5pt}{v3,r3}
\fmf{quark,tension=0,label=$q_2\uparrow$,label.side=left}{v1,v2}
\fmf{quark,tension=0,label=$q_1\downarrow$,label.side=left}{v2,v3}
\fmf{photon,label=$q\rightarrow$}{v2,r2}
\fmf{phantom}{l2,v2}
\end{fmfgraph*}} & \parbox{70pt}{\begin{fmfgraph*}(70,40)
\fmfset{curly_len}{1mm}
\fmfset{arrow_len}{2mm}
\fmfleft{l1,l2,l3}
\fmfright{r1,r2,r3}
\fmf{dbl_plain_arrow}{l1,v1}
\fmf{dbl_plain,tension=2}{v1,r1}
\fmf{dbl_plain_arrow,tension=2}{v3,v2}
\fmf{dbl_plain_arrow,tension=2}{v2,l3}
\fmf{dbl_plain,tension=2}{v3,r3}
\fmf{quark,tension=0}{v1,v3}
\fmf{photon,tension=0,left=0.3}{r2,v2}
\end{fmfgraph*}} &
\parbox{70pt}{\begin{fmfgraph*}(70,40)
\fmfset{curly_len}{1mm}
\fmfset{arrow_len}{2mm}
\fmfleft{l1,l2,l3}
\fmfright{r1,r2,r3}
\fmf{dbl_plain_arrow,tension=2}{l1,v1}
\fmf{dbl_plain}{v1,r1}
\fmf{dbl_plain_arrow,tension=2}{v3,l3}
\fmf{dbl_plain,tension=2}{v3,v2}
\fmf{dbl_plain,tension=2}{v2,r3}
\fmf{quark,tension=0}{v1,v3}
\fmf{photon,tension=0,left=0.3}{r2,v2}
\end{fmfgraph*}} \vspace*{2mm} \\
(1) & (2) & (3)
\end{tabular}
  \end{center}
\caption{Diagrams contributing to the Multi-Regge limit of the process (\ref{eq:model-DY}) in the model theory. Two diagrams where photon interacts with the opposite proton and spectator lines also should be added.}\label{fig:model-diags}
  \end{figure}

  In this limit, only diagrams in the Fig.~\ref{fig:model-diags} contribute, because ``crossed'' diagrams have at least two propagators suppressed by $\sqrt{S}$. Factors in the second and third diagrams which describe interaction of the photon with the proton or spectator line carrying large $P_1^+$ momentum can be simplified at leading power in $Q/\sqrt{S}$ as follows:
\begin{eqnarray*}
\hspace{-3mm}{\cal M}^\mu_2&\propto& e_p \bar{v}(P_1)\gamma^\mu\frac{\hat{P}_1-\hat{q}}{(P_1-q)^2}(i\lambda_{spq})\simeq e_p \bar{v}(P_1)  \frac{P_1^+ \gamma^\mu\hat{n}_-}{2(-P_1^+ q^-)}(i\lambda_{spq})= \bar{v}(P_1) (i\lambda_{spq}) \frac{i\hat{q_1}}{q_1^2} \left[ ie_p \frac{\hat{q}_1 n_-^\mu}{q_-} \right], \\
\hspace{-3mm}{\cal M}^\mu_3&\propto& e_s \frac{(2P_1 + 2q_2-q)^\mu}{(P_1+q_2)^2} \bar{v}(P_1) (i\lambda_{spq}) \simeq  e_s\frac{P_1^+ n_-^\mu}{P_1^+ q^-} \bar{v}(P_1)(i\lambda_{spq}) = \bar{v}(P_1) (i\lambda_{spq}) \frac{i\hat{q}_1}{q_1^2} \left[ -i e_s \frac{\hat{q}_1 n_-^\mu}{q_-} \right],
\end{eqnarray*}
where $\lambda_{spq}$ is the spectator-proton-quark interaction constant. Hence, ${\cal M}_2+{\cal M}_3$ is proportional to $e_p-e_s=e_q$ as well as the first diagram. When all five diagrams are taken together, the Multi-Regge limit of the amplitude can be cast into a following effective $t$-channel exchange form:
\[
{\cal M}^\mu \simeq \bar{v}(P_1) (i\lambda_{spq}) \frac{i\hat{q}_1}{q_1^2} \left(-ie_q\Gamma^\mu(q_1,q_2)\right) \frac{-i\hat{q}_2}{q_2^2} (i\lambda_{spq}) u(P_2).
\]
 This little example shows, that Fadin-Sherman vertex is independent from spins and charges of particles which are highly separated in rapidity from the photon, which is a basic prerequisite of factorization.

  In PRA we propose to modify the definition of $W^{\rm (TMD)}$ in Eq.~(\ref{eq:TMD-PM}) as:
\begin{equation}
W_{\mu\nu}^{\rm (PRA)}=\sum_{q,\bar{q}} \frac{e_q^2}{N_c} {\rm tr}\left[\Gamma_\mu(q_1,q_2) \Phi_{q}(q_1,P_1) \otimes_T \Gamma_\nu(q_1,q_2) \bar{\Phi}_{\bar{q}}(q_2,P_2)  \right].
\end{equation}
  
  This leads to the following factorization formula for the contributions of number-density TMD PDF to the structure functions~\cite{DY-paper-1}:
\begin{equation}
F_{UU}^{(1,2,\ldots)}=\sum\limits_{q,\bar{q}} \frac{e_q^2}{N_c} f_1^{q}(x_1,{\bf q}_{T1}) \otimes_T f_1^{\bar{q}}(x_2,{\bf q}_{T2})\cdot w^{(1,2,\ldots)}_{PRA}({\bf q}_{T1}, {\bf q}_{T2},Q^2),  \label{eq:PRA-F}
\end{equation}
where $x_{1,2}=Q_T e^{\pm Y}/\sqrt{S}$, $Q_T=\sqrt{Q^2+{\bf q}_T^2}$ and
\begin{eqnarray*}
w_{PRA}^{(1)}=\frac{2Q^2+{\bf q}_T^2}{2Q_T^2},\ w_{PRA}^{(2)}=\frac{({\bf q}_{T1}-{\bf q}_{T2})^2}{Q_T^2},\ w_{PRA}^{(\cos\phi)}=\sqrt{\frac{Q^2}{{\bf q}_T^2}}\frac{{\bf q}_{T1}^2-{\bf q}_{T2}^2}{Q_T^2},\ w_{PRA}^{(\cos 2\phi)}=\frac{{\bf q}_T^2}{2Q_T^2}.
\end{eqnarray*}

  Eq.~(\ref{eq:PRA-F}) is just the Eq.~(7) from our Ref.~\cite{DY-paper-1} rewritten in terms of TMD PDFs with the same normalization as in Eq.~(\ref{eq:Phi-TMD}), which is more conventional in the TMD community. This TMD PDFs are related with TMD PDFs of PRA as $f_1^{q}(x,t,\mu^2)=\Phi_q(x,t,\mu^2)/(\pi\sqrt{S} x)$, since in PRA we include the flux-factor for initial-state partons $1/(2Sx_1x_2)=1/(2Q_T^2)$ into the cross-section formula. 

  One observes, that the contributions of number-density TMD PDF to all structure functions except $F_{UU}^{(1)}$ are $Q(q_T^2/Q^2)$, as it should be, according to the TMD PM analysis, so that the only leading-power contribution to $F_{UU}^{(\cos 2\phi)}$ comes from the convolution of two Boer-Mulders functions. However, the Boer-Mulders TMD PDF is expected to be significantly smaller than number-density TMD PDF and it's effects are observable only at nonzero $q_T$. Taking into account, that values of $Q^2$ in the existing and planned experiments, such as COMPASS and NICA SPD~\cite{NICA} lie in a ballpark of $10$ GeV, the power-suppressed corrections could be important for the extraction of Boer-Mulders TMD PDF, especially in the transition region $q_T\sim Q$. 

  In Ref.~\cite{DY-paper-1} we have performed a numerical analysis with the help of Eq.~(\ref{eq:PRA-F}) and a realistic model for number-density TMD PDF, based on the Kimber-Martin-Ryskin formula~\cite{KMR} and MSTW-2008~\cite{MSTW08} set of collinear PDFs. This TMD PDF allowed us to reproduce the observed $q_T$-spectra of Drell-Yan pairs from several low-energy experiments and polarization observables measured by NuSea Collaboration~\cite{NuSea}, see Refs.~\cite{DY-paper-1, DY-paper-2} for more details. 

\begin{figure}[t]
\begin{center}
\includegraphics[width=0.49\textwidth]{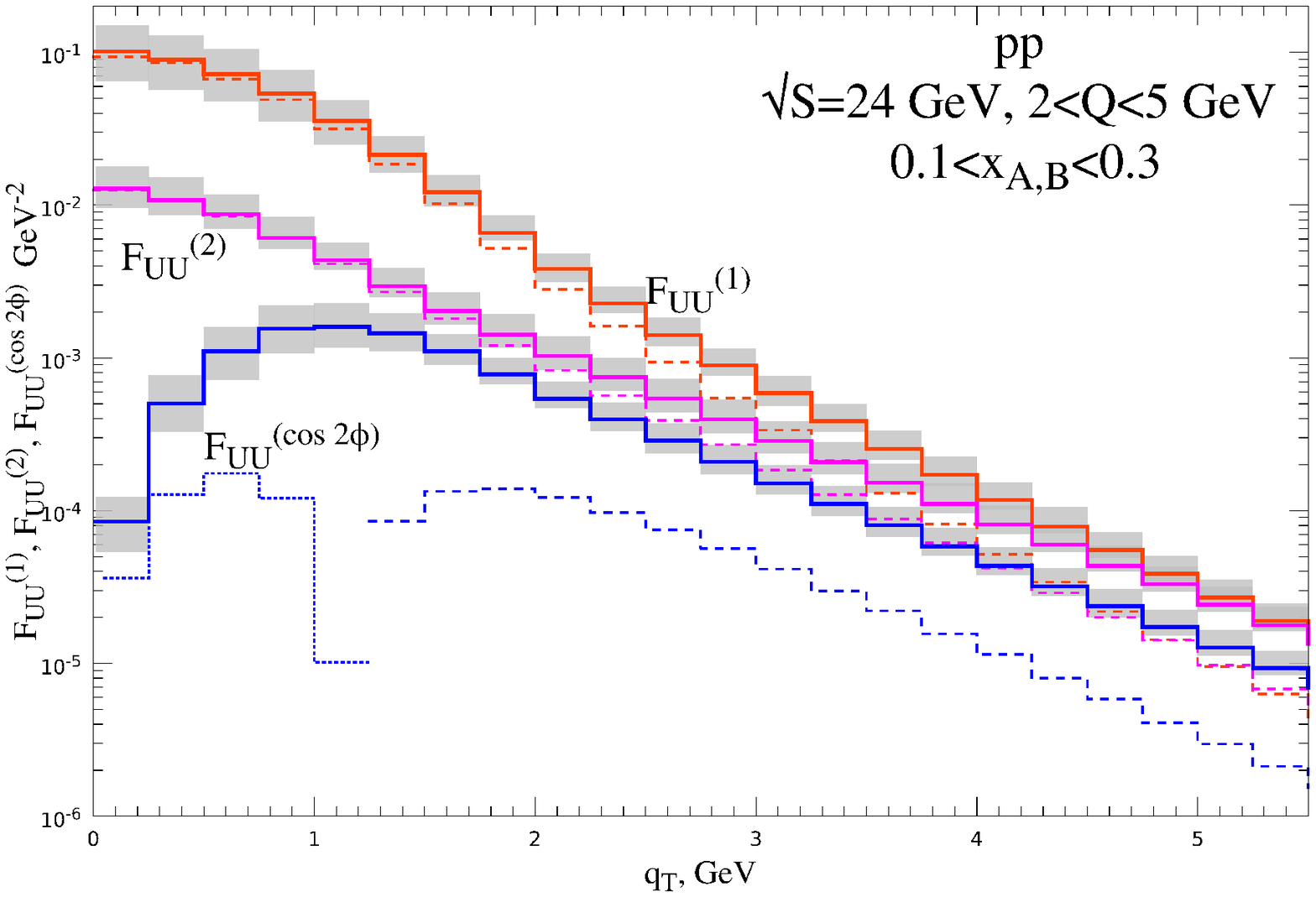}
\includegraphics[width=0.49\textwidth]{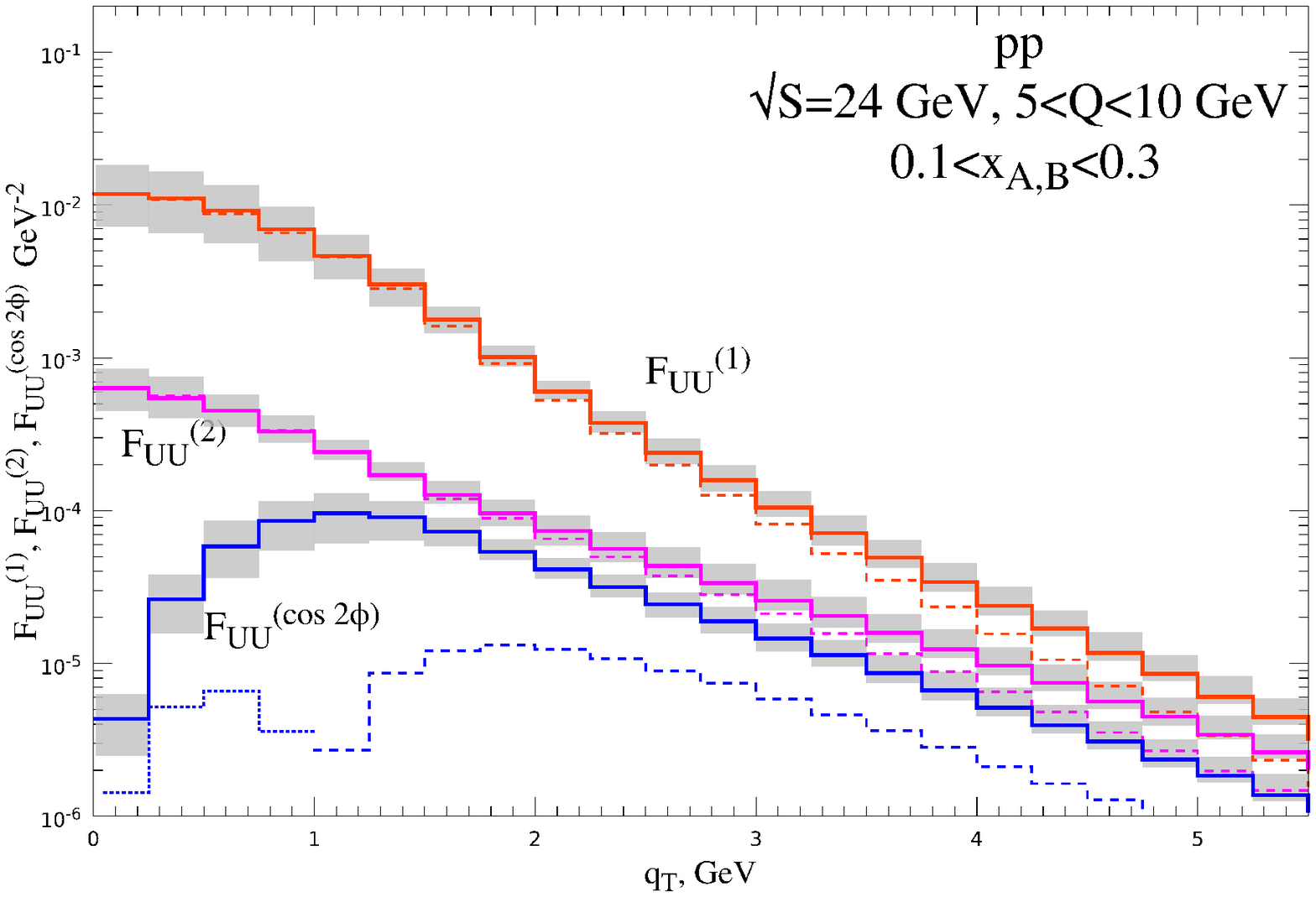}
\end{center}
\caption{Predictions for unpolarized Drell-Yan SFs $F_{UU}^{(1)}$ , $F_{UU}^{(2)}$ and $F_{UU}^{(\cos 2\phi)}$ in $pp$-collisions at $\sqrt{S} = 24$ GeV~\cite{NICA}. Solid lines with uncertainty bands -- PRA predictions. Dashed lines -- predictions in the QOS-scheme~\cite{DY-paper-1} for the default scale-choice. Short-dashed line - plot of the $(-F_{UU}^{(\cos 2\phi)})$ in the QOS scheme, since this SF in QOS scheme is negative at low $q_T$.}\label{fig:NICA-SFs}
\end{figure}

  The results for structure functions, obtained in Ref.~\cite{DY-paper-1} are shown in the Fig.~\ref{fig:NICA-SFs}. Positivity of angular distribution (\ref{eq:SFs-def}) requires $F_{UU}^{(\cos 2\phi)}\leq F_{UU}^{(1)}+F_{UU}^{(2)}$. From Fig.~\ref{fig:NICA-SFs} one can see, that $F_{UU}^{(\cos 2\phi)}$ in PRA can reach values up to a few percent of $F_{UU}^{(1)}$ at $q_T\sim 1$ GeV and up to a several tens of percent, if one increases $q_T$ closer to $Q$. This PRA results can be viewed as an estimate of a contribution of power-suppressed corrections. Measured values of $F_{UU}^{(2, \cos2\phi)}$ at this level can be interpreted as a result of power-suppressed corrections and can not serve as a clear indication of Boer-Mulders effect.

\end{fmffile}
\end{document}